\documentclass[twocolumn,a4paper,preprintnumbers,amsmath,amssymb,nofootinbib,floatfix]{revtex4}

\usepackage{color}
\usepackage{graphicx}
\usepackage{dcolumn}
\usepackage{bm}
\usepackage{latexsym}
\usepackage{epsfig}
\usepackage{rotating}
\usepackage{bm}

\newcommand{\be}{\begin{equation}}
\newcommand{\ee}{\end{equation}}
\newcommand{\beqq}{\setlength\arraycolsep{2pt}\begin{eqnarray}}
\newcommand{\eeqq}{\vspace{0cm} \end{eqnarray}}
\newcommand{\bea}{\begin{eqnarray}}
\newcommand{\eea}{\end{eqnarray}}

\begin{document}

\title{Searching for deviations from the general relativity theory with gas mass fraction of galaxy clusters  and complementary probes}

\author{R. F. L. Holanda$^{1,2,3}$} \email{holanda@uepb.edu.br}
\author{S. H. Pereira$^{4}$} \email{shpereira@feg.unesp.br}
\author{S. Santos da Costa$^{5}$} \email{simonycosta@on.br}

\affiliation{ \\$^1$Departamento de F\'{\i}sica, Universidade Estadual da Para\'{\i}ba, 58429-500, Campina Grande - PB, Brazil,
\\ $^2$Departamento de F\'{\i}sica, Universidade Federal de Campina Grande, 58429-900, Campina Grande - PB, Brazil,\\$^3$Departamento de F\'{\i}sica, Universidade Federal do Rio Grande do Norte, 59300-000, Natal - RN, Brazil.\\$^4$Universidade Estadual Paulista (Unesp)\\Faculdade de Engenharia, Guaratinguet\'a \\ Departamento de F\'isica e Qu\'imica\\ Av. Dr. Ariberto Pereira da Cunha 333\\
12516-410 -- Guaratinguet\'a, SP, Brazil
\\$^5$Observat\'orio Nacional, 20921-400, Rio de Janeiro - RJ, Brazil.}



\begin{abstract}

Nowadays, thanks to the improved precision of cosmological data, it has become possible to search for deviation from the general relativity theory with tests on large cosmic scales. Particularly, there is a class of modified gravity theories that breaks the Einstein equivalence principle (EEP) in  the electromagnetic sector,  generating variations of  the fine structure constant, violations of the cosmic distance duality relation and the evolution law of cosmic microwave background (CMB) radiation. In recent papers, this class of theories has been tested with angular diameter distances from galaxy clusters, type Ia supernovae and  CMB temperature. 

In this work we propose a new test by considering the most recent x-ray surface brightness observations of galaxy clusters jointly with type Ia supernovae and  CMB temperature. {The crucial point here is that we take into account the dependence of the x-ray gas mass fraction  of galaxy clusters on  possible variations of the fine structure constant and violations of the cosmic distance duality relation.} Our basic result is that this new approach is  competitive with the previous one and it also does not show significant deviations from general relativity.

\end{abstract}

\maketitle


\section{Introduction}

Recently, several models of modified gravity theories have been proposed in order to deal with some problems that general relativity (GR) cannot solve directly. Although it is the best gravity theory we know, GR fails when one tries to understand some local observations concerning galactic velocities in galaxy clusters or curve of rotations of spiral galaxies. The addition of a new kind of attractive matter that does not interact electromagnetically, the so-called dark matter (DM), 
is the standard solution in order to maintain GR as the background gravity theory. Nevertheless, the nature, origin and dynamics of such a new kind of matter are still a mystery.  On the other hand, recent observations of type Ia supernovae  (see \cite{suzuki} for a recent compilation and references therein) have shown that the Universe is currently accelerating and 
 in the GR context this fact can be explained only with the addition of a cosmological constant (CC) term
or with the addition of a new kind of repulsive energy, the so-called dark energy (DE) \cite{reviewDE}, whose nature is also undetermined.

However, if we take out such ingredients placed by hand, namely, DM, CC and DE,  then GR alone is not enough to explain the large amount of observational data in several scales. For this reason, alternative models have been proposed recently in order to accommodate the observations, such as massive gravity theories; 
modifications of Newtonian dynamics (MOND); 
$f(R,T)$ theories that generalize the Lagrangian of GR; 
 higher-dimension models as brane world models; 
string theories at low energies; 
and Kaluza-Klein theories, among others. Nevertheless, several of these theories explicitly break the Einstein equivalence principle (EEP) leading to  {explicit variations of some fundamental constants of nature, e.g., the fine structure constant}. This allows us to test, from an observational point of view, the degree of confidence of such modified theories.

A powerful mechanism to test the signatures of a class of modified gravity theories was developed by Hees et al. \cite{hees}. These authors showed that theories  explicitly breaking the EEP for having a nonminimal multiplicative coupling between the usual electromagnetic part of matter fields and a new scalar field (motivated by scalar-tensor theories of gravity, for instance \cite{string,klein,axion,fine,chameleon,fRL}) lead to  { variations of the fine structure constant, $\alpha$, of quantum electrodynamics} \cite{const_alpha}. Thus, the entire electromagnetic sector of the theory is also affected,  leading to a nonconservation of photon number and, consequently, to a modification of the expression of the luminosity distance, which is the basis for various cosmological evaluations. 

After that, based on the results of Hees et al., some recent papers \cite{holandaprd,holandasaulo} have also searched for deviations from GR by considering the same class of modified gravity theories that explicitly breaks the EEP ({see the next section for more details}). These studies used angular diameter distances (ADD) of galaxy clusters obtained via their x-ray surface brightness + Sunyaev-Zel'dovich effect (SZE) observations \cite{fil,bonamente}, type Ia supernovae (SNe Ia) samples \cite{suzuki} and the cosmic microwave background (CMB) temperature in different redshifts, $T_{CMB}(z)$ \cite{luzzi}. No deviation from the GR was verified, although the results do not completely rule out the models under question.

In this paper, we follow searching for deviations from GR by considering cosmological observations and the class of models that explicitly breaks the EEP in the electromagnetic sector. We show that measurements of the x-ray gas mass fraction (GMF) in galaxy clusters jointly with SNe Ia distance moduli and CMB temperature at different redshifts also furnish an interesting test for GR. In our analyses,  we use the most recent GMF sample from Ref. \cite{mantz} with galaxy clusters in the redshift range $0.078 \leq z \leq 1.063$. This sample contains 40 massive and morphologically relaxed systems obtained from the {\it Chandra} observations, with high-quality weak gravitational lensing data, fundamental to x-ray mass calibration.  We show that our analyses present competitive results with those found in Ref.\cite{holandasaulo} and no significant deviation from GR is verified. 

This paper is organized as follows: In Sec. II we briefly revise the changes on important results from the standard cosmology  when a  {time variation of the fine structure constant is assumed}. In Sec. III we present GMF measurements that can be used to test these kinds of modifications. The  samples and analyses are in Sec. IV. The Section V  shows the results and in Sec. VI we conclude.

\section{Break of EEP}

\subsection{Deviations from the standard relations  }

Some kinds of modified gravity theories that explicitly break the EEP can be represented by a matter action that has the following form:
\begin{equation}
S_{m}=\sum_i \int d^4x\sqrt{-g}h_i(\phi)\mathcal{L}_i(g_{\mu\nu},\Psi_i)\,,\label{action}
\end{equation}
where $\mathcal{L}_i$ are the Lagrangians for different matter fields $\Psi_i$ and $h_i(\phi)$ represents a nonminimal coupling between the extra field $\phi$ (motivated by several alternative models \cite{string,klein,axion,fine,chameleon,fRL}) and $\Psi_i$. When $h_i(\phi)=1$ we recover the standard GR scenery. When the coupling is with the electromagnetic sector of the Lagrangian, the fine structure constant $\alpha$ of the quantum electrodynamics turns out to be time dependent \cite{fine,const_alpha}, $\alpha \propto h^{-1}(\phi(t))$. The main implication of a such  time dependence is that the photon number is not conserved along geodesics,  leading to a modification of the luminosity distance expression, $D_L$, and also to the violation of the so-called cosmic distance-duality relation (CDDR), which happens to be written as \cite{hees}:
\begin{equation}
\frac{D_L(z)}{D_A(z)(1+z)^2}=\eta (z)= \sqrt{\frac{h( \phi_0 )}{h( \phi (z))}}\,,\label{DLDA}
\end{equation}
where $D_A$ is the angular diameter distance \cite{distance} and $\eta(z)$ parametrizes the deviation from standard GR, which corresponds to $\eta(z)=1$ and can be recovered for the present time $[ \phi_0 \equiv \phi (z = 0)]$. Some usual parametrizations for $\eta(z)$ are \cite{holanda2010,holandasaulo}: 
\begin{itemize}
\item P1: $\eta(z)=1+\eta_0 z$

\item P2: $\eta(z)=1+\eta_0 z/(1+z)$

\item P3: $\eta(z)=(1+z)^{\eta_0}$

\item P4: $\eta(z)=1+ \eta_0 \ln(1+z)$
\end{itemize}
where $\eta_0$ is the parameter to be constrained. The limit $\eta_0=0$ corresponds to the standard GR results. In terms of the parametrization $\eta(z)$, a possible modification of the fine structure constant  {[$\alpha(z)=\zeta(z)\alpha_0$]} can be written as \cite{hees}:

\begin{equation}
\label{alpha}
\frac{\Delta \alpha}{\alpha}= \zeta(z) -1 = \eta^2 (z) -1.
\end{equation}

Moreover, also due to the nonconservation of the photon number, a variation of the  evolution  of the  CMB radiation is also expected, which affect its temperature distribution as \cite{hees}:
\begin{equation}
T(z)=T_0(1+z)[0.88+0.12 \eta^2(z)]. \label{T}
\end{equation}

{The relation (\ref{alpha}) allows one to transform constraints on $\Delta \alpha/\alpha$ into constraints on $\eta(z)$ and inversely. In this way, the authors of Ref. \cite{hees} used a laboratory constraint  on $\Delta \alpha/\alpha$ \cite{sri}, and a limit on $\eta_0$ in the above functions P1, P2, P3 and P4 was obtained, such as $\eta_0= 10 \pm 14 \mbox{x} 10^{-8}$. However, although very impressive, it relies on only one observation at $z=0$. From the $\Delta \alpha/\alpha$ values of 154 absorbers observed with the VLT \cite{kin} and 128 absorbers observed by the Keck observatory \cite{mur} in the redshift range $0.2 \leq z \leq 0.42$, the $\eta_0$  parameter was found as $\approx 10^{-6}$. On the other hand, different signs for $\Delta \alpha/\alpha$ (and $\eta_0$) were obtained through analyses by using the two data sets separately (see Table IV in \cite{hees}). Finally, limits on $\eta(z) $ also can be obtained by considering the CMB data ($z=1100$). The {\it Planck} Collaboration found: $\Delta \alpha /\alpha \approx 10^{-3}$ \cite{adea}. However, this limit is inferred  in the flat $\Lambda$CDM model, considering purely adiabatic initial conditions with an almost scale-invariant power spectrum and no primordial gravity waves, and is weakened by opening up the parameter space to variations of the number of relativistic species or the helium abundance (see Figs. 5 and 6 in \cite{adea}). Particularly, the Hubble constant value shifts from $H_0=67.3 \pm 1.2$ km/s/Mpc (with $\alpha$ constant) to $H_0=65.1 \pm 1.8$ km/s/Mpc (with $\alpha$ varying), exacerbating the tension with the value of the Hubble constant found in \cite{riess}: $H_0=73.8 \pm 2.4$ km/s/Mpc  using cepheids and SNe Ia. In this way, it is still worthwhile to search for variations of fundamental constants by using other observations, preferably independent of the cosmological model.}

 {Moreover, it is important to point out that in recent years several papers have tested the CDDR's validity by using  different astronomical quantities, such as angular diameter distances from galaxy clusters plus SNe Ia, SNe Ia plus $H(z)$ data, gas mass fraction of galaxy clusters plus SNe Ia, gamma ray burst plus $H(z)$ data, CMB temperature, SNe Ia plus baryon acoustic oscillations, and gravitational lensing plus SNe Ia. As a result, the validity of the CDDR was verified at least within a 2$\sigma$ C.L. (see the table in \cite{hba2016} for  recent results). Particularly, some tests use the following galaxy cluster data:  angular diameter distance via their Sunyaev-Zel'dovich effect  + x-ray observations \cite{distance2} and the x-ray gas mass fraction measurement \cite{gon,wang2013}. However, it was shown very recently that these quantities are  strongly dependent on not only the CDDR validity, but also on the fine structure constant \cite{hac2016,hls2016}. This way, by taking into account the  direct relation between variations of $\alpha$ and violations of the CDDR in the SZE + x-ray technique [Eq. (3)], the authors of Refs. \cite{holandaprd,holandasaulo} used  angular diameter distances from galaxy clusters, luminosity distances from SNe Ia, and $T_{CMB}(z)$ and $\eta(z)$ functions to search for cosmological signatures of the EEP  {breaking}. No significant indication of violation of the standard framework was obtained (see Table I in \cite{holandasaulo}).}

\begin{figure*}[t]
\centering
\includegraphics[width=0.47\textwidth]{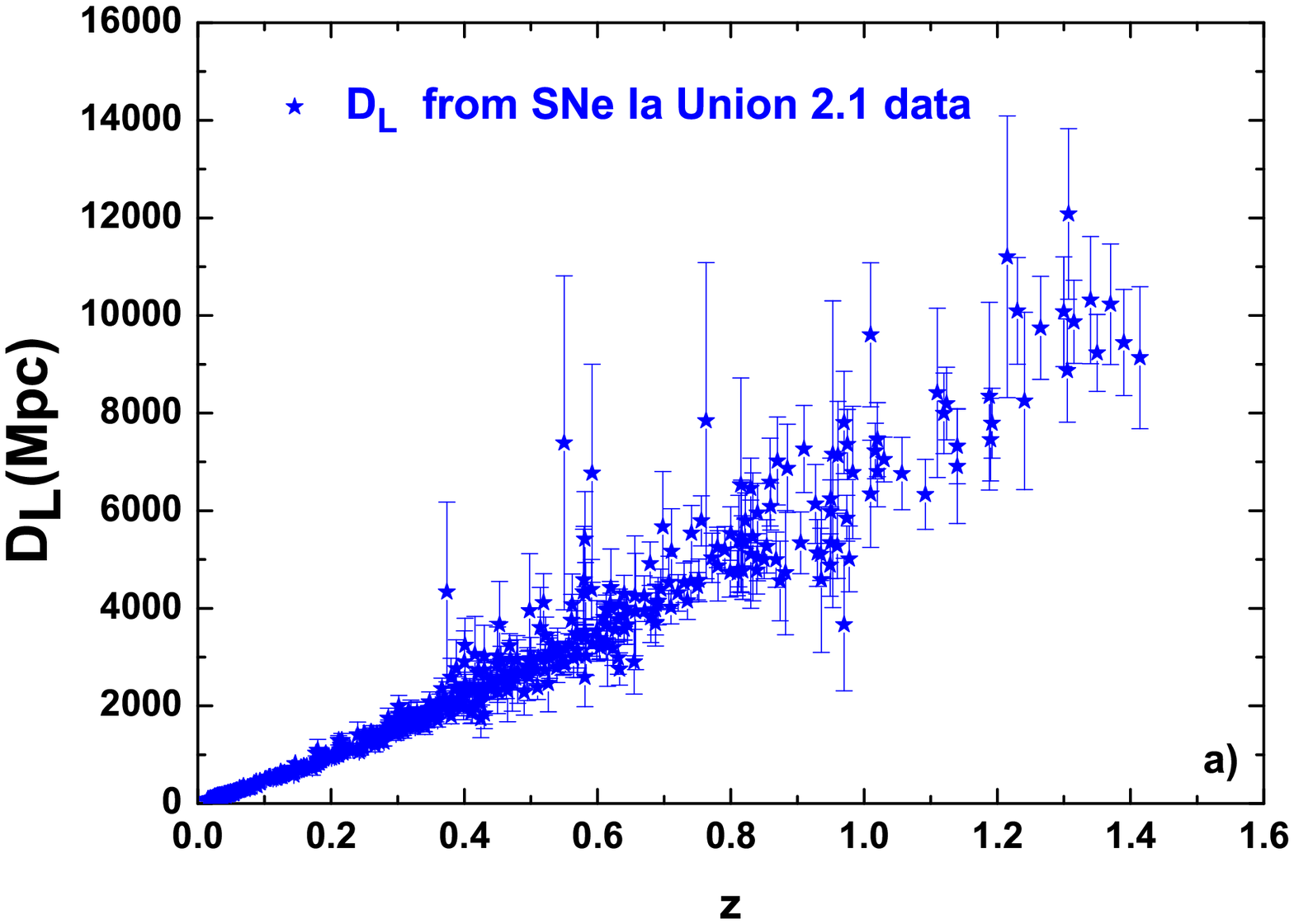}
\hspace{0.3cm}
\includegraphics[width=0.47\textwidth]{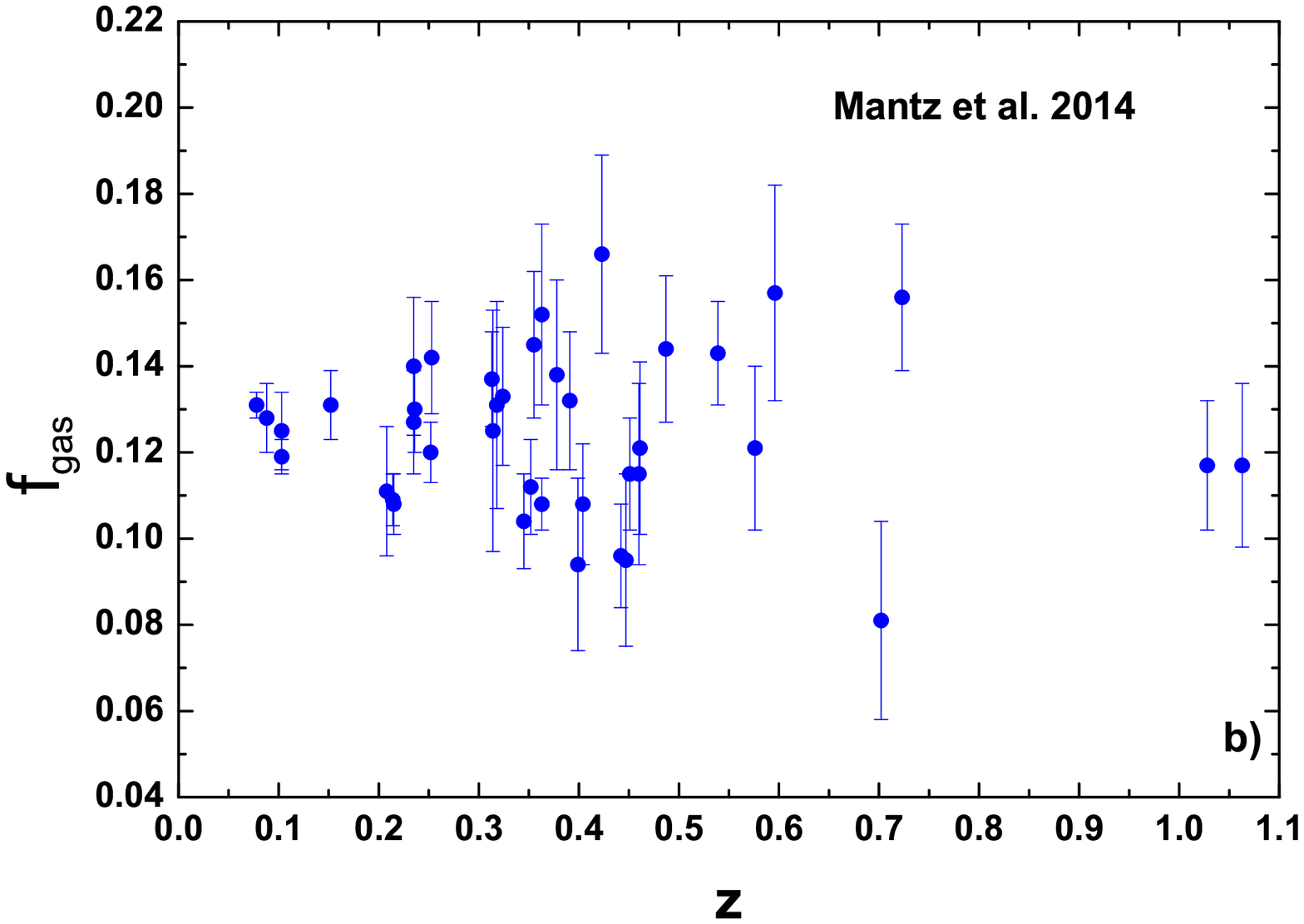}
\hspace{0.3cm}
\includegraphics[width=0.47\textwidth]{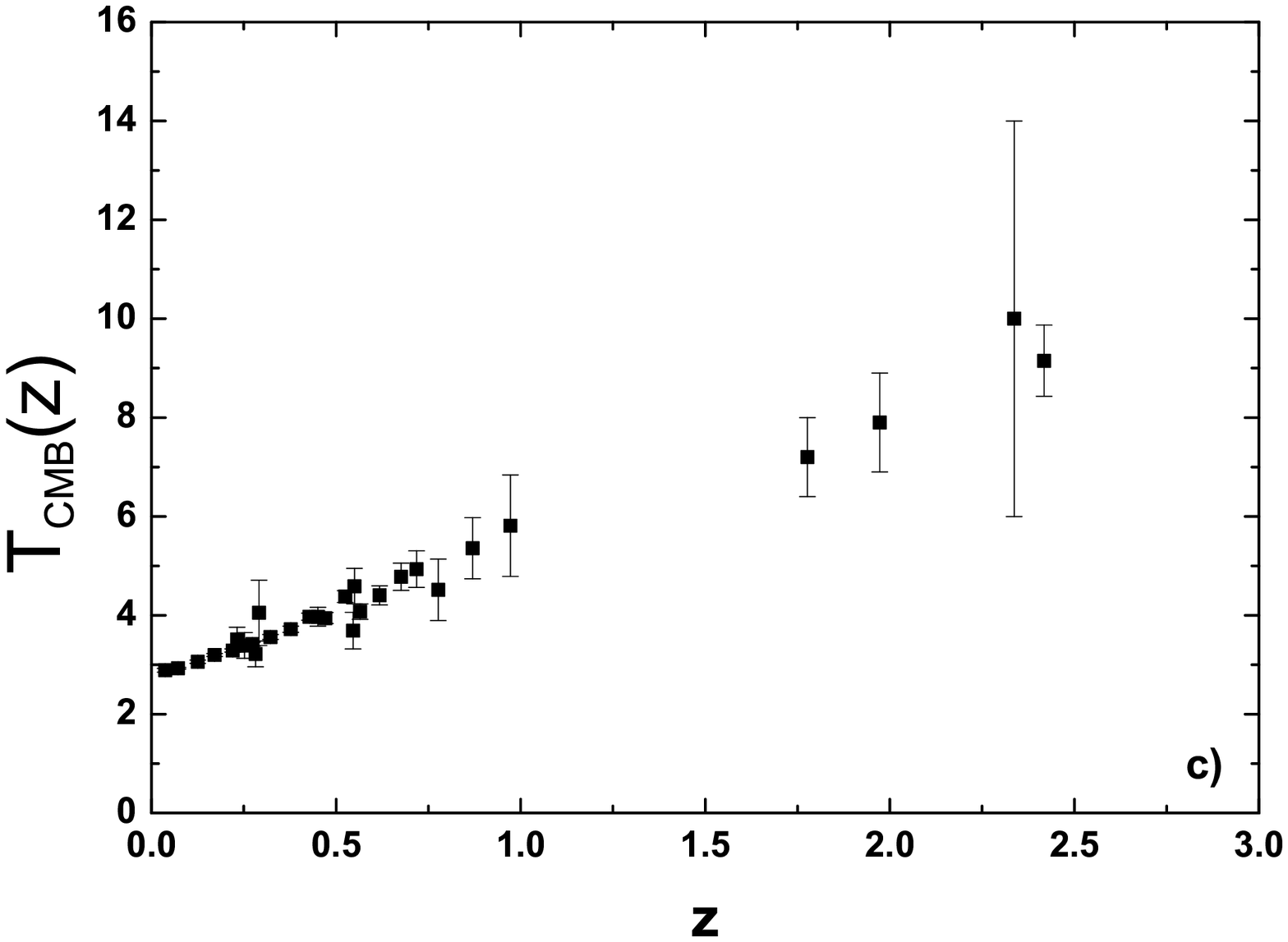}
\caption{ Plots of (a) 580 distance moduli of SNe Ia \cite{suzuki}; (b) 40 GMF data \cite{mantz}; and (c) 36 $T_{CMB}(z)$ data \cite{luzzi}. }
\end{figure*}

In this work, we search for the EEP breaking by testing jointly the CDDR (\ref{DLDA}) by using GMF + SNe Ia observations and the CMB temperature evolution  law (\ref{T}). The  {decisive} point here is that we take into account the dependence of the x-ray GMF with  $\zeta(z)$ and $\eta(z)$ \cite{hls2016}.

\subsection{Consequences for gas mass fraction measurements}

Here we discuss the consequences of the EEP breaking on GMF measurements and explain the basic equations used in our analyses.

The baryonic matter content of galaxy clusters is dominated by an x-ray emitting intracluster gas via predominantly thermal bremsstrahlung (see \cite{sar} for more details). On the other hand, the total mass within a given radius $R$  is obtained by assuming that the intracluster gas is in hydrostatic equilibrium. Thus, the gas mass fraction is defined as \cite{ale}:

\begin{equation}
 f_{gas}=\frac{M_{gas}}{M_{tot}},
  \label{eq3.14}
\end{equation}
where $M_{tot}$ is the total mass  { (dominated by dark matter)}  and  $M_{gas}$ is the gas mass obtained by  integrating the gas density model.  { Since these structures are the largest virialized objects in the Universe, the cluster baryon fraction is a faithful representation of the cosmological average baryon fraction $\Omega_b/\Omega_M$, in which $\Omega_b$ and $\Omega_M$ are, respectively, the fractional mass density of baryons and all matter. In this way, one may expect $f_{gas}$  to be same at all $z$ \cite{ale}. Thus, x-ray observations of galaxy clusters can  be used to constrain cosmological parameters from the following expression \cite{ale}:}
\begin{equation}
\label{GasFrac}
f^{obs}_{X-ray}(z)=N\left[\frac{D_L^* D_A^{*1/2}}{D_L D_A^{1/2}}\right],
\end{equation}
 {where the symbol * denotes quantities from a fiducial cosmological model  used in the observations (usually a flat $\Lambda$CDM model where $\eta=1$). The parameter $N$ defines the astrophysical model of the cluster, such as stellar mass fraction, nonthermal pressure and the depletion parameter {\footnote{This indicates the amount of  baryons that are thermalized within the cluster potential \cite{ale} and it tends to be lower than unity because not all the accreting shock-heated baryons relax within the dark matter halo \cite{pla,bat}.}}.{ Multiplying by the ratio in brackets computes the expected measured gas fraction $f^{obs}_{X-ray}$ when the observer incorrectly assumes a $\Lambda$CDM
cosmology while the actual cosmology is different from that, with distances $D_L$ and $D_A$, and $\eta(z)$ not necessarily equal to unity. This
rescaling allows a direct comparison to literature measurements of the cluster gas fraction.}} 

 {Our method is completely based on the recent results from Refs.\cite{gon,hls2016}. In Ref.\cite{gon}, the authors showed that the gas mass fraction measurements  extracted from x-ray data are affected by a possible departure of $\eta =1$ and Eq. (\ref{GasFrac}) must be rewritten as}
\begin{eqnarray}
\label{GasFrac3}
f^{obs}_{X-ray}(z) &=& N \left[\frac{\eta(z)^{1/2}D_L^{*3/2}}{D_{L}^{3/2}}\right].
\end{eqnarray}
 {The $\eta$ parameter appears after using (\ref{DLDA}) in  {the} denominator}.  On the other hand, in Ref.\cite{hls2016} it was shown that the gas mass fraction measurements  extracted from x-ray data also are affected by a possible departure of $\zeta(z)=1$, such  {that} 
\begin{equation}
f_{X-ray} \propto  [\zeta(z)]^{-3/2}.
\end{equation}  
 {Thus, by considering the class of theories explored by Hees et al. \cite{hees}, from Eq. (3):}
\begin{equation}
f_{X-ray} \propto  \eta(z)^{-3}.
\end{equation}
As one may see, the quantity $f^{obs}_{X-ray}$ may still deviate from  its true value by a factor $\eta(z)^{-3}$, which does not have a counterpart on the right side in Eq. (\ref{GasFrac3}). Thus, this expression has to be modified to 
\begin{eqnarray}
\label{GasFrac4}
f^{obs}_{X-ray}(z) &=& N \left[\frac{\eta(z)^{7/2}D_L^{*3/2}}{D_L^{3/2}}\right].
\end{eqnarray}
 {So}, it is possible to obtain the luminosity distance of a galaxy cluster from its GMF by:
\begin{eqnarray}
\label{dl}
D_L &=& \eta(z)^{7/3}D_L^{*}[N/f^{obs}_{X-ray}(z)]^{2/3}.
\end{eqnarray}

Finally, we define the distance module of a galaxy cluster as:
\begin{equation}
 \mu_{cluster}(\eta ,z)=5\log[\eta(z)^{7/3}D_L^{*}[N/f^{obs}_{X-ray}(z)]^{2/3}]+25, 
\end{equation} 
 { where $D_L^*$ is in Mpc.} As one may see, if we have SNe Ia distance module measurements, $\mu(z)$, at identical redshifts of galaxy clusters, we can put observational constraints on the $\eta(z)$ function.

\begin{figure*}[htb]
\centering
\includegraphics[width=0.47\textwidth]{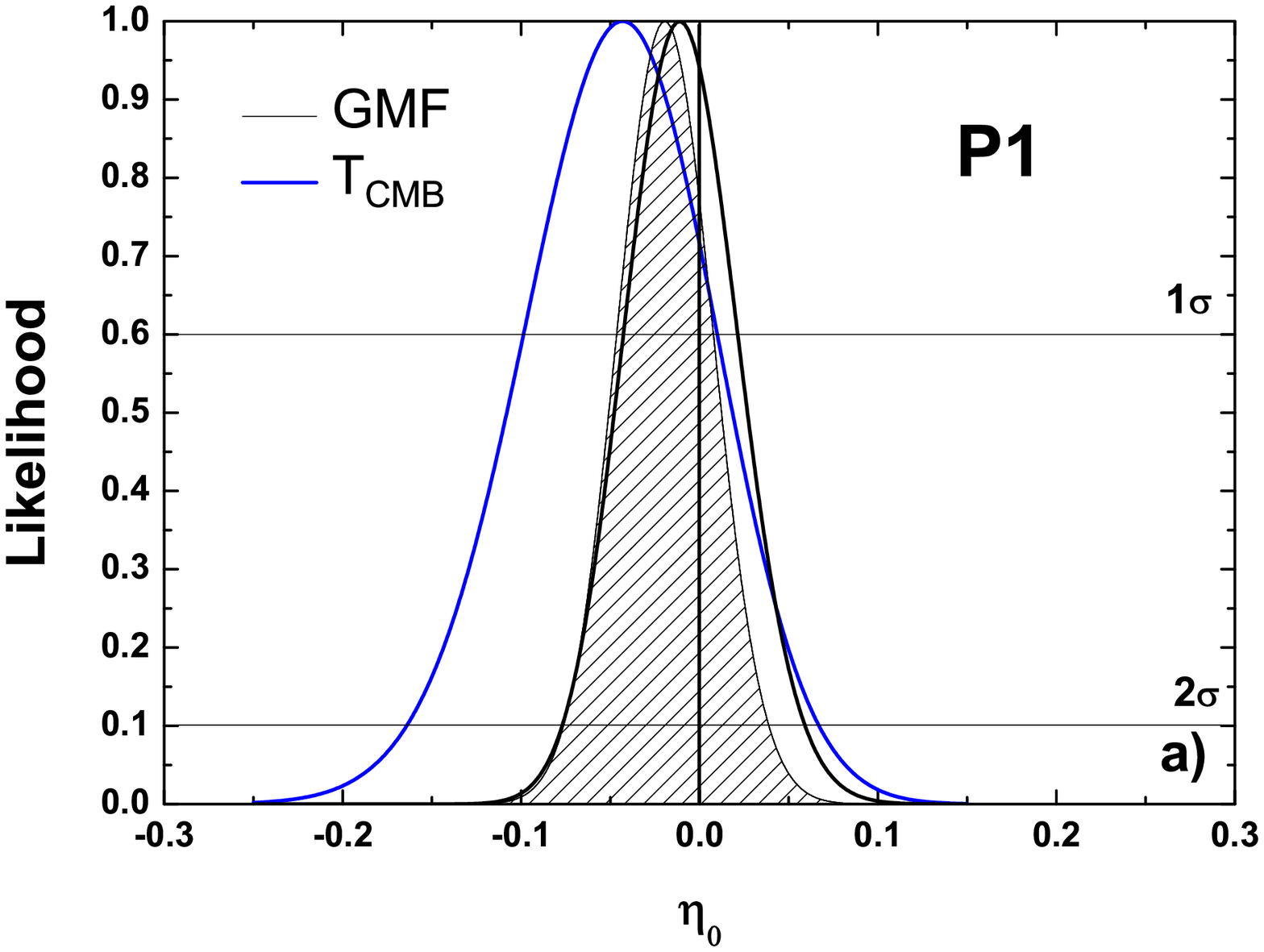}
\hspace{0.3cm}
\includegraphics[width=0.47\textwidth]{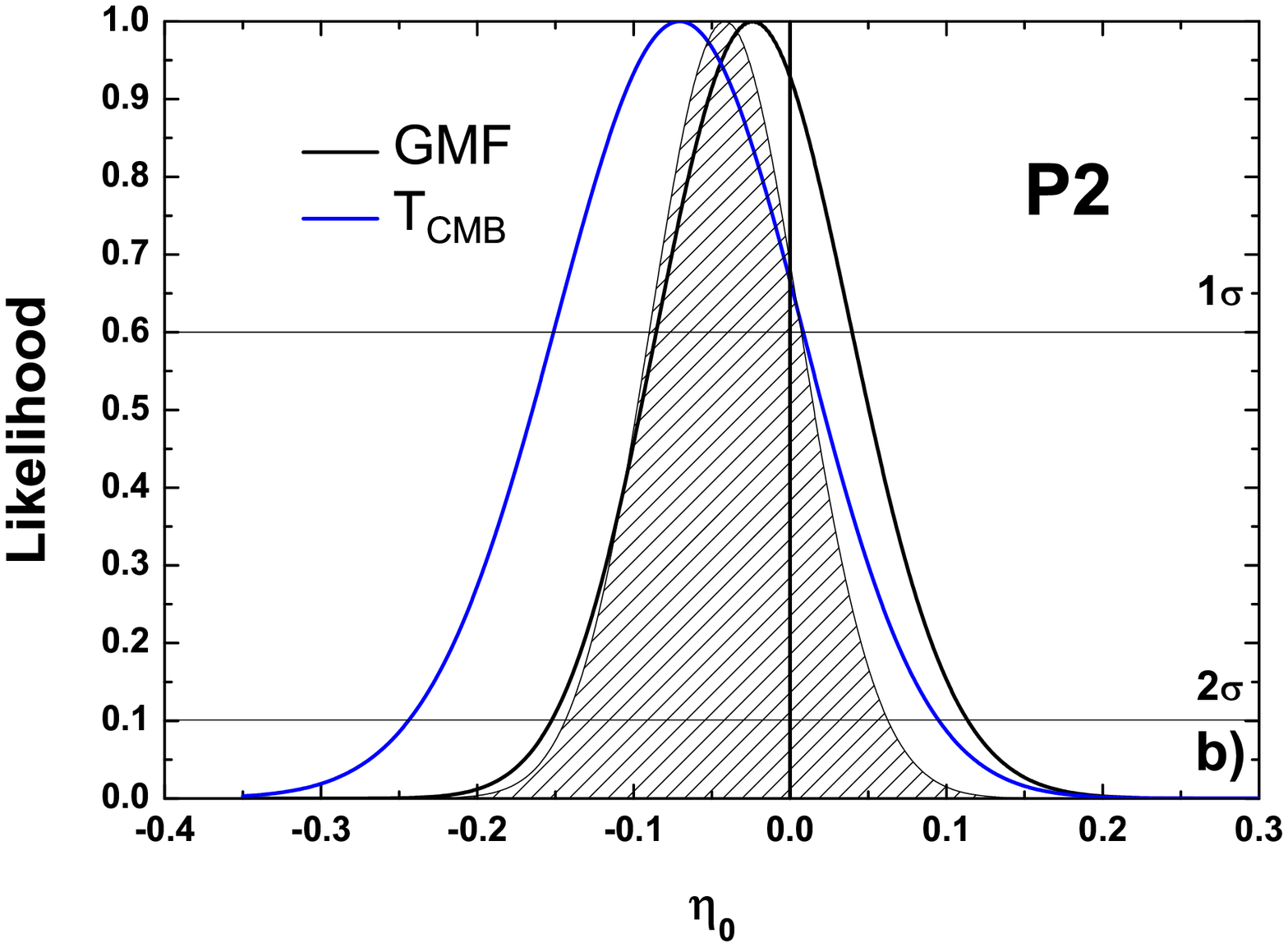}
\caption{In both figures, the solid black and blue lines correspond to analyses by using SNe Ia + GMF and  $T_{CMB}(z)$, respectively. The dashed area corresponds to the joint analysis (SNe Ia + GMF + $T_{CMB}(z)$). We plot the results by using the (a) parametrization P1 and (b) P2.}
\end{figure*}

\section{Data}

In our analyses, we use the following data set:

\begin{itemize}
\item The full SNe Ia sample is formed by 580 SNe Ia data compiled by Suzuki et al. \cite{suzuki}, the so-called Union2.1 compilation, with redshift range $0.015 \leq z \leq 1.4$ [see Fig. 1(a)].  The Union2.1 SNe Ia compilation is an update of the Union2 compilation and all SNe  Ia were fitted using SALT2 \cite{guy2007}. All analyses and cuts were developed in a blind manner, i.e., with the cosmology hidden.  In order to perform our test we need SNe Ia and galaxy clusters in the identical redshifts. Thus, for each GC in the GFM sample, we select SNe Ia with redshifts obeying the criteria $|z_{GC} - z_{SNe}| \leq 0.006$ and calculate the following weighted average for the SNe Ia data:
\begin{equation}
\begin{array}{l}
\bar{\mu}=\frac{\sum\left(\mu_{i}/\sigma^2_{\mu_{i}}\right)}{\sum1/\sigma^2_{\mu_{i}}} ,\hspace{0.5cm}
\sigma^2_{\bar{\mu}}=\frac{1}{\sum1/\sigma^2_{\mu_{i}}}.
\end{array}\label{eq:dlsigdl}
\end{equation}
Following  \cite{suzuki} we added a 0.15 systematic error to SNe Ia data.

\item 40  GMF measurements in redshift range $0.078 \leq z \leq 1.063$ from Mantz et al. \cite{mantz}. This is the most recent GMF sample [see Fig. 1(b)]. Differently from previous GMF measurements, these authors measured the GMFs in spherical shells at radii near $r_{2500}$\footnote{This radius is that one within which the mean cluster density is 2500 times the critical density of the Universe at the cluster's redshift.}, rather than  integrated at all radii ($< r_{2500}$). This reduces significantly the corresponding theoretical uncertainty in gas depletion from hydrodynamic simulations (see Fig. 8 in their paper). Moreover, the bias in the mass measurements from x-ray data arising by assuming hydrostatic equilibrium was calibrated by robust mass estimates for the target clusters from weak gravitational lensing (see also \cite{app}).

\item  { The $T_{CMB}(z)$ sample is composed by 37 points in redshifts between 0 and 2.5 (see Fig. 1c). The 36 data points ($z\neq 0$)  are based on different observations, among them: the Sunyaev-Zeldovich effect, the rotational excitation of CO lines, the fine structure of carbon atoms, and X-ray data from galaxy clusters \cite{luzzi}. We also use the estimation of the current CMB temperature $T_0 = 2.725 \pm 0.002$ K  from the CMB spectrum as estimated from the COBE collaboration \cite{mather}.}
\end{itemize}

\section{Analyses and Results}

\begin{table*}[ht]
\caption{A summary of the current constraints on the parameters $\eta_0$ for P1, P2, P3 and P4,  from angular diameter distance from galaxy clusters and different SNe Ia samples. The symbol * corresponds to the angular diameter distance (ADD) from Ref. \cite{fil} and ** to the angular diameter distance from Ref. \cite{bonamente}.}
\label{tables1}
\par
\begin{center}
\begin{tabular}{|c||c|c|c|c|c|}
\hline\hline Reference & Data sample &$\eta_0$ (P1)& $\eta_0$ (P2)& $\eta_0$ (P3)& $\eta_0$ (P4)
\\ \hline\hline 
\cite{holandaprd} & $ADD^{*}$   + SNe Ia      & $0.069 \pm 0.106$    & $0.000 \pm 0.135$ & - &-\\
\cite{holandasaulo} & $ADD^{**}$   + SNe Ia  + $T_{CMB}$ & $-0.005 \pm 0.025$& $-0.048 \pm 0.053$ & $-0.005\pm 0.04 $&$-0.005 \pm 0.045$ \\
\cite{holandasaulo} & $ADD^{*}$   + SNe Ia  + $T_{CMB}$  & $-0.005 \pm 0.032$& $-0.007 \pm 0.036$ &$ 0.015 \pm 0.045$ & $0.015 \pm 0.047$\\
\bf{This paper} & GMF + SNe Ia  + $T_{CMB}$  & $-0.020 \pm 0.027$ & $-0.041 \pm 0.042$  & -  & - \\

\hline\hline
\end{tabular}
\end{center}
\end{table*}

We evaluate our statistical analysis by defining the likelihood distribution function, ${\cal{L}} \propto e^{-\chi^{2}/2}$, where 
\begin{eqnarray}
\chi^2 &=& \sum_{i=1}^{40}\frac{(\bar{\mu}(z_i)-\mu_{cluster}(\eta,z_i))^2}{\sigma_{obs}^2}\nonumber \\ &&+ \sum_{i = 1}^{37}\frac{{\left[ T(z_i) - T_{i,obs} \right] }^{2}}{\sigma^{2}_{T_i, obs}} ,
\label{chi}
\end{eqnarray} 
with $\sigma_{obs}^2= \sigma^2_{\bar{\mu}} + \sigma^2_{\mu cluster}$ and  $T(z)$  given by Eq.(4). In our analyses, the normalization factor $N$  is taken as a nuisance parameter so that we marginalize over it. The EEP breaking is sought for allowing deviations from $\eta=1$, such as: (P1) $\eta(z)=1+\eta_0 z$ and (P2) $\eta(z)=1+\eta_0 z/(1+z)$. As one may see, if $\eta_0=0$ from the data analyses, then GR is verified.  
 
Our results are plotted in Figs. 2(a) and 2(b) for each parametrization. Note that in each case the  solid blue and black lines correspond to analyses by using separately  CMB temperature and GMF + SNe Ia data in Eq.(\ref{chi}), respectively. The dashed areas are the results from the joint analysis, i.e., the complete Eq. {\eqref{chi}} with CMB temperature + GMF + SNe Ia. In Table I we put our 1$\sigma$ results from the joint analyses for each parametrization and several $\eta_0$ values already present in the literature which consider correctly possible variations of $\alpha$ and $\eta$ in their analyses.  As one may see, our results are in full agreement with each other and with the previous one regardless of the galaxy cluster observations and  $\eta(z)$ functions used. Moreover, our analyses present competitive results with those found in Ref.\cite{holandasaulo} and no significant deviation from GR is verified.

\section{Conclusion}

In this work we continue the analyses in order to test some cosmological signatures of modified gravity theories that explicitly break the EEP. As already pointed out by Hees et al. \cite{hees}, alternative theories described by the action (\ref{action}) lead naturally to variations of the standard constants of physics, modifications of the cosmic distance duality relation and also the CMB temperature distribution. 
The possibility of testing such theories is of great interest. 

The main contribution of this work is to use the dependence of current measurements of x-ray gas mass fraction in galaxy clusters on the EEP validity to test the class of theories pointed out in \cite{hees}. Following the analyses of previous works \cite{holandaprd,holandasaulo}, where ADD + SNe Ia + $T_{CMB}$ data were used to constrain possible deviations from GR, here we consider GMF + SNe Ia + $T_{CMB}$ data in order to obtain constraints on the $\eta_0$ parameter for the two parametrizations P1 and P2 of CDDR. As presented in Table I, no significant deviation from GR was found.

\begin{acknowledgements}
R. F. L. H. acknowledges financial support from  {Conselho Nacional de Desenvolvimento Cient\'ifico e Tecnol\'ogico} (CNPq) and UEPB (Grants No. 478524/2013-7 and No. 303734/2014-0). S. H. P. is grateful to CNPq, for financial support, Grants No. 304297/2015-1.  {S. S. C. acknowledges financial support from Coordena\c{c}\~{a}o de Aperfei\c{c}oamento de Pessoal de N\'ivel Superior (CAPES). The authors thank the anonymous referee for the valuable suggestions.}
\end{acknowledgements}

\end{document}